\documentclass[twocolumn,twoside,slac_two]{revtex4}
\usepackage{graphicx}
\usepackage{fancyhdr}
\def\jp{J/\psi}
\def\pp{\psi'}
\def\dsb{{\bar D}^{*0}}
\def\lsim{\raisebox{-0.6ex}{$\stackrel{\textstyle <}{\sim}$}}

\def\beq{\begin{equation}}
\def\eeq{\end{equation}}

\begin{document}
\title{Molecular Quarkonium}

%

\author{M.~B.~Voloshin}
\affiliation{FTPI, University of Minnesota, 116 Church St.~SE, Minneapolis, MN
55455, USA and \\ITEP, B.~Cheremushkinskaya 25, 117218 Moscow, Russia}

\begin{abstract}
I discuss topics related to four-quark states of the `molecular quarkonium'
type, i.e. resonances that could be considered as (dominantly) made from a heavy
meson and antimeson. Of the so far observed resonances such picture is very
likely applicable to the state X(3872), and I also discuss its possible
relevance to the peak near the $D^* {\bar D}^*$ threshold in $e^+e^-$
annihilation.
\end{abstract}

\maketitle

\thispagestyle{fancy}


\section{Introduction}
It has been pointed out long time ago~\cite{vo} that there should exist bound
and/or resonant states of heavy hadrons, dubbed `molecular' and arising due to
the strong force between the light constituents of the heavy hadrons. Indeed,
considering the interaction between a heavy meson, consisting of a heavy quark
$Q$ and a light antiquark ${\bar q}$, and an antimeson (${\bar Q} q$), it is
clear that there is a normal strong force between such objects arising from an
exchange of a light ${\bar q}q$ pair, which can be thought of as an exchange of
light mesons. The range and the strength of the resulting strong interaction
does not depend on the mass $m_Q$ of the heavy quark $Q$ in the limit $m_Q \to
\infty$. If the interaction corresponds to an attraction in some states of the
heavy meson-antimeson pair, it should inevitably give rise to bound states in
those channels at sufficiently large $m_Q$. However, given the lack of detailed
knowledge of the strong interaction between heavy hadrons, it has not been
clear, whether the charmed quark is heavy enough for the reasoning based on the
heavy quark limit to be applicable, and thus whether resonances of such type
with hidden charm do exist. Another difficulty of identifying states with hidden
charm as molecular, or more generically as four-quark systems, is that in most
models of light meson exchange between charmed mesons~\cite{vo,nat,ek} the
interaction with the strongest attraction arises in channels with zero isospin
and with non-exotic quantum numbers $J^{PC}$, which can be also realized as a
conventional $c {\bar c}$ charmonium. For this reason it could be a typical
situation that some dominantly four-quark states may be indistinguishable from
states of charmonium, unless some other properties would favor a molecular
interpretation. One such interpretation was argued in the past~\cite{drgg} for
the resonance in the $e^+e^-$ annihilation at the threshold of the $D^* {\bar
D}^*$ production, then called $\psi(4028)$, and presently referred to~\cite{pdg}
as $\psi(4040)$. It was noted that the cross section for production of the $D^*
{\bar D}^*$ pairs accounted for more than one half of the total production of
charmed meson pairs in spite of the very strong suppression by the $P$ wave
kinematical factor $p^3$ at energy close to the threshold. Thus it was suggested
that there is a resonance very close to the threshold actually made from $D^*
{\bar D}^*$ meson pairs.

The idea of molecular states was revived in a modern context by the
observation~\cite{belle1} in the Belle experiment of the resonance X(3872)
decaying into $\pi^+ \pi^- \jp$, and produced in the decays $B \to X \, K$ of
the $B$ mesons. The resonance is quite narrow, and its width is not yet resolved
($\Gamma_X < 2.3 \,$MeV at 90\% CL), and the updated value of its mass~\cite{hm}
$m_X=(3871.2 \pm 0.6)\,$MeV coincides within the errors with the mass of the
$D^0 \dsb$ meson pair: $M(D^0)+ M(D^{*0})=3871.2 \pm 1\,$MeV. The extreme
closeness of the newly observed resonance to the meson pair threshold is very
suggestive of an interpretation~\cite{cp,ps,mv1,nat2} of this resonance as a
`molecule' dominantly consisting of neutral charmed meson pairs: $D^0 \dsb \pm
D^{*0} {\bar D}^0$.

\section{X(3872)}
The discovery~\cite{belle1} and observation~\cite{cdf,d0,babar1} of X(3872)
through the decay $X \to \pi^+ \pi^- \jp$ was followed by finding~\cite{belle2}
also the decays $X  \to \pi^+ \pi^- \pi^0 \jp$ and $X \to \gamma \jp$. Clearly,
the co-existence of the latter decays with the discovery mode $X \to \pi^+ \pi^-
\jp$ implies an isospin breaking in X(3872). Indeed the two-pion state and the
three-pion one have opposite G parity, and one or the other would have to be
forbidden if the isospin symmetry were applicable in the decays of $X(3872)$.
Furthermore, the existence of the radiative decay $X \to \gamma \jp$ requires
$X$ to have positive C parity, and the two-pion system emitted in the decay $X
\to \pi^+ \pi^- \jp$ has to be in a C-odd state, which corresponds to the
isospin of this system equal to one: $I_{\pi\pi}=1$. Thus the $X(3872)$
resonance is required to have a substantial $I=1$ component, which is impossible
for a pure $c {\bar c}$ charmonium. In other words, the resonance has to contain
light $u {\bar u} $ and/or $d {\bar d}$ quark pairs in addition to hidden charm,
which thus qualifies it as a four-quark state. Moreover, the decays $X \to \pi^+
\pi^- \jp$ and $X  \to \pi^+ \pi^- \pi^0 \jp$ have comparable
rates~\cite{belle2}, which implies that the X(3872) resonance does not have a
definite isospin, and is mixture of $I=1$ and $I=0$ states. In addition, a
subsequent study~\cite{belle3} of the angular correlations and the dipion mass
spectrum in the decay $X \to \pi^+ \pi^- \jp$ strongly favors the spin-parity
assignment $J=1^+$ for the X(3872). Combined with the extreme proximity of its
mass to $M(D^0)+ M(D^{*0})$ the known properties of X(3872) most strongly
suggest that a sizeable part of the wave function of this resonance is
contributed by the C-even $S$-wave state of charmed meson pair $D^0 \dsb +
D^{*0} {\bar D}^0$, which thus has the quantum numbers $J^{PC}=1^{++}$.

The binding energy $w$ of the neutral meson pair in X(3872) is small: likely
$w\, \lsim \,1\,$MeV. In this scale the mass gap $\delta$ between the
corresponding threshold for the pair of charged mesons and that for the neutral
ones $\delta= M(D^+)+ M(D^{*+})- M(D^0)- M(D^{*0}) \approx 8\,$MeV is large
enough to suppress the weight of the state $D^+ {\bar D}^{*-} + D^- {\bar
D}^{*-}$ in the wave function of X(3872), thus explaining the unusual isotopic
properties of the resonance.

Generally, the wave function of the resonance X(3872) can be written in terms of
a Fock decomposition:
\begin{eqnarray}
|X \rangle &=& a_0 \, |D^0 {\bar D}^{0*} + D^{0*} {\bar D}^{0} \rangle
\nonumber \\
&+& a_1 \, |D^+ {\bar D}^{-*} + D^{+*} {\bar D}^{-} \rangle + \sum \, | {\rm
other } \rangle~,
\label{fock}
\end{eqnarray}
where first two terms correspond to the `molecular' states of charmed meson
pairs, and the ``other" states in the sum describe the admixture of
non-molecular states, such as e.g. a $^3P_1$ charmonium $c {\bar c}$ pair. The
latter states are localized at shorter distances within the range of the strong
interaction, while the `molecular' part refers to the motion of the charmed
mesons at (mostly) the distances beyond the range of strong
interaction~\cite{mv1,mv2}. At the latter distances the coordinate wave function
$\phi(r)$ for each pair is that of a free motion:
\beq
\phi_n(r)=\sqrt{\kappa_n \over 2\pi} \, {\exp (-\kappa_n \, r) \over r}~,
\label{phin}
\eeq
for the meson combination $(D^0 {\bar D}^{0*} + D^{0*} {\bar D}^{0})/\sqrt{2}$
and
\beq
\phi_c(r)=\sqrt{\kappa_c \over 2\pi} \, {\exp (-\kappa_c \, r) \over r}~,
\label{phic}
\eeq
for $(D^+ {\bar D}^{-*} + D^{+*} {\bar D}^{-})/\sqrt{2}$. The virtual momenta
$\kappa_n$ and $\kappa_c$ are related to the binding energy $w$ and the reduced
mass of the two-meson system $m_r \approx 970\,$MeV as $\kappa_n = \sqrt{2
\, m_r \, w} \approx 44\,$MeV\,$\sqrt{w/{\rm MeV}}$ and ${\kappa_c} = \sqrt{2
m_r \, (\delta+w)} \approx 125 \,$MeV.

The dominance of the $D^0 {\bar D}^{*0} + D^{*0}{\bar D}^0$ at long distances
translates into a substantial isospin violation in the processes determined by
the `peripheral' dynamics, examples of which are apparently the observed decays
$X \to \pi^+  \pi^- J/\psi$ and  $X \to \pi^+  \pi^- \pi^0 J/\psi$. It is quite
likely however that this isospin-breaking behavior is only a result of the
`accidentally' large mass difference $\delta \approx 8 \,$MeV between $D^+
D^{*-}$ and $D^{*0} {\bar D}^{*0}$. Therefore
it is natural to expect that at shorter distances within the range of the strong
interaction the isospin symmetry is restored and at those distances the wave
function of $X(3872)$ is dominated by $I=0$, so that the `core' of the wave
function of X is an isotopic scalar. Then the wave functions of the $D^+
D^{*-}+D^-D^{*+}$ and  $D^0 {\bar D}^{*0} + D^{*0}{\bar D}^0$ should coincide at
short distances, which allows to estimate the relative statistical weight of
these two states in the X(3872) resonance (i.e. the ratio of the corresponding
coefficients in Eq.~\ref{fock}):
\beq
\lambda \equiv {|a_1|^2 \over |a_0|^2} =
{\kappa_n \over {\kappa_c}}~.
\label{wrat}
\eeq

Strictly speaking, the wave functions in Eq.~\ref{phin} and Eq.~\ref{phic}
cannot be
applied at short distances in the region of strong interaction, where the mesons
overlap with each other and cannot be considered as individual particles. In
order to take into account this behavior an `ultraviolet' cutoff should be
introduced. One widely used method for introducing such cutoff is to consider
the meson wave functions only down to a finite distance $r_0$, at which distance
the boundary condition of the state being that with $I=0$ is imposed (a
discussion of a similar situation can be found e.g. in Ref.\cite{mv3}). An
alternative, somewhat more gradual cutoff, described by parameter $\Lambda$, can
be introduced\cite{suzuki} by subtracting from the wave functions (\ref{phin})
and (\ref{phic}) an expression $c \, e^{-\Lambda r}/r$.
One can also readily see, that an introduction of any such cutoff eliminates
relatively more of the charged meson wave function than of the neutral one, thus
reducing the estimate of the relative statistical weight as compared to that in
Eq.~\ref{wrat}, so that Eq.~\ref{wrat} gives in fact the upper bound for the
ratio.

\section{Peripheral Decays of X(3872)\,: $X \to D^0 {\bar D}^0 \pi^0$ and $X \to
D^0 {\bar D}^0 \gamma$}
\begin{figure}[h]
\centering
\includegraphics[width=80mm]{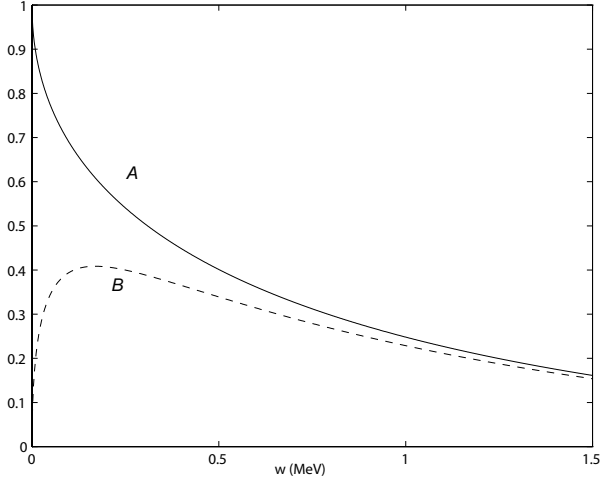}
\caption{The dependence of the coefficients $A$ and $B$ in Eq.~\ref{gampi} on
the binding energy $w$.} \label{gampw}
\end{figure}

The peripheral  $D^0 {\bar D}^{*0} + D^{*0}{\bar D}^0$ component with the
quasi-free mesons should give rise to the decays $X \to D^0 {\bar D}^0 \pi^0$
and $X \to D^0 {\bar D}^0 \gamma$ due to the underlying decays of the $D^{*0}$
and $\dsb$ mesons. The widths of the corresponding underlying decays can be
deduced from the available data~\cite{pdg}: $\Gamma_\pi \equiv \Gamma(D^{*0} \to
D^0 \pi^0)= 43 \pm 10 \,$KeV and $\Gamma_\gamma \equiv \Gamma(D^{*0} \to
D^0\gamma) = 26\pm 6\,$KeV. In the C-even state X(3872) the amplitudes of the
decays $D^{*0} \to D^0 \pi^0$ and $\dsb \to {\bar D}^0 \pi^0$ interfere
constructively and the result for the rate of the decay of $X$ can be written
as~\cite{mv1}
\beq
\Gamma(X \to D^0 {\bar D}^0 \pi^0)=2 |a_0|^2 \, \Gamma_\pi \, \left [ A(w)
+ \eta \, B(w) \right ]~,
\label{gampi}
\eeq
with $A(w)$ describing the non coherent contribution of the decays from $D^{*0}$
and $\dsb$ and $B(w)$ being the interference term. The dependence of these
coefficients on the binding energy $w$ is shown in Figure~\ref{gampw}. As can be
seen from the plot the interference is quite important at $w$ as small as
0.1\,MeV and its relative importance increases with $w$, since at stronger
binding the distance between the mesons in X(3872) becomes shorter. The overall
fall-off of both $A$ and $B$ at larger $w$ is the result of the decrease of the
phase space for the decay.

\begin{figure}[h]
\centering
\includegraphics[width=80mm]{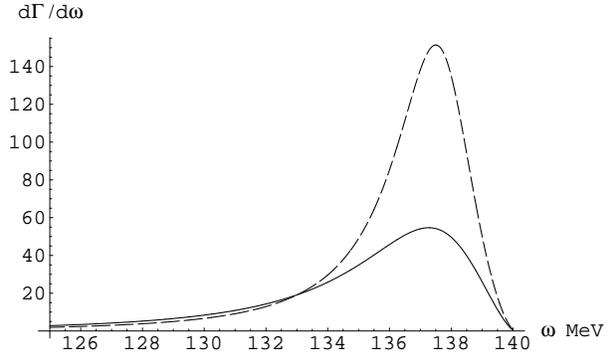}
\caption{The photon energy spectrum in the peripheral decay $X(3872) \to D^0
{\bar D}^0 \gamma$, at $w=0.3\,$MeV (dashed) and at $w=1\,$MeV (solid)}
\label{gspectrum}
\end{figure}

The amplitudes of the radiative transitions $D^{*0} \to D^0 \gamma$ and $\dsb
\to {\bar D}^0 \gamma$ interfere destructively in the decay $X(3872) \to D^0
{\bar D}^0 \gamma$.~\footnote{Such effect of destructive interference in the
photon emission is known is subradiance~\cite{abd, dicke}.} The expression for
the rate has the form~\cite{mv1,mv2}:
\beq
\Gamma(X \to D^0 {\bar D}^0 \gamma)=\Gamma_\gamma \, |a_0|^2 \left
[ 1-{2 \kappa_n \over \omega} \, \arctan {\omega \over 2 \kappa_n} \right
]~,
\label{gamg}
\eeq
where the photon energy $\omega$ can be approximated by that in the free meson
decay $D^{*0} \to D^0 \gamma$: $\omega \approx \omega_0 = 137\,$MeV, due to the
fact that the weak binding only slightly spreads the photon spectrum in the
decay $X \to D^0 {\bar D}^0 \gamma$ near $\omega_0$, as can be seen from the
plots shown in Figure \ref{gspectrum}. The shape of the spectrum is determined
by the wave function of the meson pair in the X(3872) and thus a measurement of
this shape is a direct probe of the motion of the mesons in the `molecule'.

It can be noted that the other kinematically allowed radiative decay channel $X
\to D^+ D^- \gamma$ should be quite suppressed~\cite{mv2}. The suppression of
the direct peripheral contribution due to the transitions $D^{*\pm} \to D^\pm
\gamma$ arises from the discussed small overall statistical weight of the $D^+
{\bar D}^{-*} + D^{+*} {\bar D}^{-}$ state in the wave function (\ref{fock}) and
from a strong destructive interference between the amplitudes of the processes
$D^{*+} \to D^+ \gamma$ and $D^{*-} \to D^- \gamma$ due to a significantly
shorter spatial separation of the charged mesons in X(3872). The contribution
due to the final-state rescattering $X \to D^0 D^0 \gamma \to D^+ D^- \gamma$ is
also very small~\cite{mv2} due to the $p^3$ behavior of the scattering  $D^0 D^0
\to D^+ D^-$ very near the threshold below the $\psi(3770)$ resonance. The most
efficient way of producing the $D^+ D^- \gamma$ final state in the decay of
X(3872) appears to be due to the transition $X(3872) \to \psi(3770) \gamma$ from
the `core' of the X resonance. Notice however that such transitions result in
the photon energy $\omega \approx 100\,$MeV i.e. well below the peak from the
peripheral process (cf. Figure \ref{gspectrum}).

\section{The Core of X(3872)}

\subsection{Spin Selection Rule}
The structure of the `core' of the X(3872) resonance represented by the ``other"
states in the wave function (\ref{fock}) is largely unknown and in discussing
this part we have to resort to general considerations and quite approximate
estimates. One such general property based on the heavy quark limit for charmed
quarks is the spin-selection rule for the $c {\bar c}$ pair in
X(3872)~\cite{mv4}. Namely, unlike a generic four quark state, the $S$ wave
C-even $(D^0 {\bar D}^{*0}+ {\bar D}^0  D^{*0})$ state is unique in the sense
that the total spin of the heavy $c {\bar c}$ pair is fixed: $S_{c {\bar c}}=1$.
Indeed, for the S wave there is no orbital motion, and the total spin of the
state is given by the vector sum of the spins of the $c {\bar c}$ and $u {\bar
u}$ quark pairs: ${\vec J}= {\vec S}_{c {\bar c}}+ {\vec S}_{u {\bar u}}$.
However the states of the type ${ S}_{c {\bar c}}=0\, \otimes \, { S}_{u {\bar
u}}=1$ and ${ S}_{c {\bar c}}=1\, \otimes \, { S}_{u {\bar u}}=0$ have negative
C parity: C=-1.
Only the state $ {S}_{c {\bar c}}=1\, \otimes \, { S}_{u {\bar u}}=1$ results in
C=+1. The total spin of the light quark pair is not `traceable', i.e. can be
flipped in the mixing with the ``core" states. The heavy quark spin however is
conserved in the limit $m_Q \to \infty$.
Therefore one arrives at the spin selection rule (valid up to
$O(\Lambda_{QCD}/m_c)$): the states with $S_{c {\bar c}}=1$ dominate in the Fock
sum in Eq.~\ref{fock} for X(3872).

As one application of the spin selection rule one can consider hadronic
transitions from X(3872) to the charmonium states. The spin selection rule
implies that the transitions to the spin-singlet charmonium levels such as
$X(3872) \to \pi \pi \eta_c$ should be suppressed while the transitions to
spin-triplet states should be allowed. The latter include the observed decays $X
\to \pi^+ \pi^- \jp$, $X \to \pi^+ \pi^- \pi^0 \jp$ and $X \to \gamma \jp$, and
also the yet unobserved decays to the spin-triplet $\chi_{cJ}$ levels: $X \to
\pi^0 \chi_{cJ}$. The conservation of the spin of the $c {\bar c}$ pair implies
the statistical behavior of the decay rates:
\beq
\Gamma(X \to \pi^0 \chi_{cJ}) \propto (2J+1)~,
\eeq
and an approximate estimate of the absolute rate~\cite{mv4} suggests that the
width of the decay $X \to \pi^0 \chi_{c1}$ can be as large as about $0.3 \,
\Gamma(X \to \pi^+ \pi^- \jp)$.

\subsection{Production of X(3872)}
The core of the wave function of the X(3872) resonance is located at
short distances. Thus it is natural to expect that the production of the
resonance in hard processes proceeds through the production of the core.
Furthermore, the quantum numbers $J^{PC}=1^{++}$ can be realized in a
$^3P_1$ $c {\bar c}$ state, so that one can expect that the production
properties of the X resonance should be similar to those of charmonium
with the rates scaled by the statistical weight of the core $|a_c|^2$.
The hard processes where X(3872) has been observed so far are the decays
$B \to X K$ and the production in the proton-antiproton annihilation.
The analysis by CDF~\cite{cdf} suggests that the production of the
X(3872) in the $p {\bar p}$ collisions is indeed similar to that of the
$\pp$ charmonium resonance. The similarity between the production of
X(3872) and of the pure charmonium states allows to approximately
estimate the weight factor of the core $|a_c|^2$. Indeed, it is an
experimental fact that the known charmonium states are produced in $B$
decays in association with a single Kaon with approximately the same
rate (within a factor of two). One might expect then that the core
states of the $X(3872)$ are produced in similar decays at approximately
the same rate, so that the only suppression factor in the observed~\cite{belle1}
rate of decays $B \to X \,K$ is that of the
probability weight $|a_c|^2$, i.e. few percent.

Furthermore, as discussed, the core part of the wave function (\ref{fock}) is
expected to be dominantly an isotopic scalar, thus within such production
mechanism one should expect that the decays $B^0 \to X \, K^0$ and $B^+ \to X \,
K^+$ should have approximately equal rates~\cite{mv4}. This conclusion is
significantly different from the calculation~\cite{bk} of the ratio of these
decay rates, based on the mechanism where the resonance is produced through its
peripheral meson component. The latter mechanism predicts ${\cal B}(B^0 \to X \,
K^0) \ll {\cal B}(B^+ \to X \, K^+)$. The current data~\cite{babar2} on these
decays
\beq
{{\cal B}(B^0 \to X \, K^0) \over {\cal B}(B^+ \to X \, K^+)}=0.50 \pm 0.30 \pm
0.05
\eeq
tend to favor the core production mechanism, although an improvement of the
experimental accuracy would be certainly helpful.

\section{Models}
The finding of X(3872) and the realization that this is likely a dominantly
molecular-type state were quite unexpected, and the problem of understanding the
dynamics of such states and predicting other possible instances of `molecules'
still remains. Although early papers~\cite{vo,nat,ek} have provided the general
idea of molecular hadronic systems, a detailed description of the strong
interaction involving heavy and light quarks is still to be developed. In this
section I briefly review some models which have been suggested after the
discovery of the X(3872).

The model suggested by Swanson~\cite{es} involves a calculation of the dynamics
of coupled channels including the $D {\bar D}^{*}$, ${\bar D} D^{*}$ pairs and
also the pairs $\rho \jp$ and $\omega \jp$, assuming a certain specific form of
the Hamiltonian for rearrangement of quarks between the hadrons. The model
produced a successful prediction of the decay $X \to \omega \jp$ ($X \to \pi^+
\pi^- \pi^0 \jp$) with the rate comparable to that of the discovery decay mode
$X \to \rho \jp$ ($X \to \pi^+ \pi^- \jp$). The model provides a detailed
description of the decays of X(3872) to these channels as well as to $D {\bar D}
\pi$, $D {\bar D} \gamma$, and $\pi^0 \gamma \jp$. I am not aware of predictions
of this model for other resonances of this type with hidden charm, however the
model predicts a $1^{++}$ resonance in the $B {\bar B}^{*}$ system at 10562\,MeV
and a $0^{-+}$ state at 10545\,MeV.

In the scheme of Maiani {\it et.al.}~\cite{mppr} the resonance X(3872) belongs
to a large class of four-quark states for which the building blocks are
color-antisymmetric diquarks and antidiquarks. In this model this resonance is
identified as a $1^{++}$ state of the structure $X_u = [cu]_{S=1} \times [{\bar
c} {\bar u}]_{S=0}+ 
[cu]_{S=0} \times [{\bar c} {\bar u}]_{S=1}$. The model predicts also
$X_d=[cd][{\bar c}{\bar d}]$ resonance with the mass splitting
$M(X_d)-M(X_{(u)}) \approx 8 \pm 3\,$MeV, and also charged partners $X^\pm$.
None of the predicted resonances with masses very close to that of X(3872) were
found so far. It can be mentioned in connection with a possible resonance of the
$X_d$ type, that even if it exists, it may be quite difficult to observe it,
since it is likely to be broad due to the allowed decay into pairs of neutral
mesons: $D^0 \dsb$ and $D^{*0} {\bar D}$. This remark however does not apply to
the charged states $X^\pm$.

\begin{figure*}[t]
\centering
\includegraphics[width=135mm]{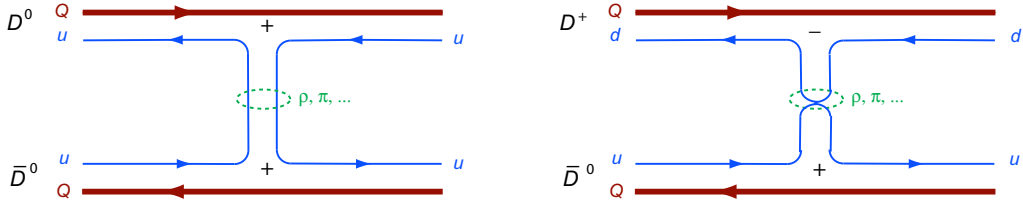}
\caption{The interaction through an isovector exchange between the neutral
mesons has opposite sign compared to that between a neutral and a charged meson}
\label{force2}
\end{figure*}

Karliner and Lipkin\cite{kl} have recently considered a nonrelativistic quark
model with pairwise interactions between the constituent quarks and antiquarks
described by an oscillator potential and proportional to the gluon-exchange-type
color correlation within each pair. The Hamiltonian in the model has the form
\beq
H=\sum_i{p_i^2 \over 2 m_i}+ \sum_{i \neq j} {V_0 \over 4} \cdot \lambda_c^i
\cdot \lambda_c^j \cdot r^2_{ij}~,
\label{klh}
\eeq
with $\lambda_c$ being the color $\lambda$ matrices. The authors find that the
model predicts four-quark bound states of the $b {\bar c} q {\bar q}$ type, but
not of the $c {\bar c} q {\bar q}$ one. They suggest that the resonance X(3872)
can still be explained within this model, once the spin-dependent forces are
taken into account.

The scheme considered by Braaten and Kusunoki~\cite{bk2} phenomenologically
treats the near-threshold dynamics of the $D^0 \dsb + {\bar D} D^{*0} $ pair as
being exclusively and universally determined by the pole in the scattering
amplitude, corresponding to X(3872), and conversely, the properties of the
resonance as being totally determined by the meson pairs. This approach
corresponds to an `extreme' molecular treatment of the resonance. In other
words, it assumes the limit, where any other states in the Fock decomposition in
Eq.~\ref{fock} are neglected in comparison with the first term, whose coordinate
wave function is given by Eq.~\ref{phin}. Clearly, such approach would miss the
properties of X(3872) determined by its `core', in particular, as discussed
above, it appears that the core dominates the production of X in hard processes.

As a general remark, it can be mentioned that existence of the X(3872) as a
dominantly $D^0 \dsb + {\bar D} D^{*0} $ shallow bound state, does not
necessarily imply existence of nearby resonances, in particular of the charged
partners $X^\pm$. Indeed, if a strong force due to exchange of isovector
light-meson states ($\rho, \pi, \ldots$) contributes to the binding in the $D^0
\dsb + {\bar D} D^{*0} $ channel, the sign of such force would be reversed in
the charged states, as can be readily seen from the Figure~\ref{force2}: the
relative sign of the two vertices for the exchanged isovector state is different
between the two shown systems.

\section{$D^* {\bar D}^*$ Threshold in $e^+e^-$ Annihilation}
The very strong enhancement of the $e^+e^-$ production of the $D^* {\bar D}^*$
meson pairs very near the threshold is known for a long time, and it served as
motivation for the suggestion~\cite{drgg} that this is an effect of a
`molecular' type $P$-wave $J^{PC}=1^{--}$ state of $D^* {\bar D}^*$. Indeed
according to the new CLEO-c measurements~\cite{poling} (which also agree with
the old data~\cite{pdg}) the cross section for $e^+e^- \to D^* {\bar D}^*$ at
$E_{c.m.} = 4030\,$ correspond in units of $R$ to $R_{D^* {\bar D}^*} \approx
0.75$ even though the velocity $v$ of the $D^*$ mesons at this energy is $v
\approx 0.1$ and value of $R_{D^* {\bar D}^*}$ is proportional to the $P$ wave
factor $v^3 \approx 10^{-3}$. It appears quite unlikely that such a strong
enhancement of the production is possible without a singularity near the $D^*
{\bar D}^*$ threshold. Furthermore, the features in the cross section of the
$e^+e^-$ annihilation into channels with pairs of lighter mesons: $D {\bar D}$,
$D^* {\bar D}$ and $D_s {\bar D}_s$ around the same energy, although present,
are still of moderate amplitude~\cite{poling}.  This behavior points to an
existence of a near threshold resonance, which is dominantly coupled to the $D^*
{\bar D}^*$ channel and has only moderate coupling to pairs of lighter mesons.

It should be mentioned that the strong $p^3$ growth of the cross section for
production of the $D^* {\bar D}^*$ pairs above the threshold leads to the
behavior where the local maximum of either the total cross section of the
$e^+e^-$ annihilation (located close to 4040\,MeV, hence the notation
$\psi(4040)$~\cite{pdg}) or the cross section for the $D^* {\bar D}^*$
production, generally does not coincide with the actual position of the
resonance. This position should be below the maximum of the $D^* {\bar D}^*$
production cross section, and can be either below, or above the threshold, or
even between the thresholds for the neutral $D^{*0} {\bar D}^{*0}$ and charged
$D^{+}*  D^{*-}$ vector mesons, which thresholds are split by about 6.6\,MeV.

The possible existence of a resonance dominantly coupled to $D^* {\bar D}^*$
generally leads to interesting features in the cross section for production of
lighter meson pairs due to channel coupling~\cite{mv5}. A detailed experimental
study of these features can potentially provide a very interesting information
on the details of strong interaction between charmed mesons, and might shed some
new light on understanding `molecular'-type states. The phenomenon of reflection
of the threshold onset in one channel on another coupled channel is well known
(see e.g. in the textbook \cite{ll}, Sect. 147). However the specifics of the
pure $P$ wave dynamics in the meson production amplitudes in the $e^+e^-$
annihilation and the apparent presence of resonance near the threshold bring in
interesting features, which justify a separate study of this phenomenon in the
discussed context.

In order to illustrate the reflection of a resonance near the $D^* {\bar D}^*$
threshold we neglect here the isotopic mass splitting between the $D^*$ mesons
and note that the resonance couples to a specific coherent mixture of the two
possible production $P$-wave amplitudes for the $D^* {\bar D}^*$ channel: one
with the total spin of the vector mesons $S=0$ and the other with
$S=2$.~\footnote{Another generally allowed amplitude, corresponding to $S=2$ and
an $F$-wave, has to be small near the threshold and is neglected here.} Thus the
final state of the $D^* {\bar D}^*$ pair can be considered as one channel,
referred here as $H$ (heavy). A channel with a lighter meson pair, i.e. $D {\bar
D}$, $D^* {\bar D}$, or $D_s {\bar D}_s$, is denoted here as $L$ (light). The
c.m. momentum in either of the $L$ channels is already large and varies only
slightly across the considered narrow energy range near the $D^* {\bar D}^*$
threshold, and this variation is neglected here.

Let $W$ denote the total c.m. energy $E$ in the $e^+e^-$ annihilation relative
to the threshold of production of the two vector mesons, each with the mass $M$:
$W = E - 2M$, and let the `nominal' position of the resonance be at the complex
energy corresponding to $W = W_0 - i \Gamma_0/2$. According to the general
quantum-mechanical consideration, based on the unitarity and analyticity of the
production and scattering amplitudes (see e.g. in the textbook \cite{ll}, Sects.
133 and 145) at energy above the `heavy' threshold, the resonant production
amplitude $A_H$ has the form
\beq
A_H={b_H \, k \over W - W_0 + {i \over 2} \, ( \Gamma_0 + g_H^2 \, k^3)}~~~,
\label{ah}
\eeq
and the production amplitudes for each of the light channels are parametrized as
\beq
A_L=a_L + {b_L  \over W - W_0 + {i \over 2} \, ( \Gamma_0 + g_H^2 \, k^3)}~~~,
\label{al}
\eeq
where $k = \sqrt{M W}$ is the c.m. momentum of the $D^* {\bar D}^*$ pair. The
coefficients $a_L, b_L,$ and $b_H$ are complex parameters, which are also taken
to be constant in the present consideration. The magnitudes and the complex
phases of these coefficients result from the rescattering among the `light'
channels and the corresponding absorptive parts associated with these channels.
It is worth noting that there is no reason to expect the phase of the
coefficient $b_H$ to vanish at $k=0$, due to the existence of `light' inelastic
channels strongly coupled to the `heavy' meson pair. The normalization
convention used here for the amplitudes corresponds to the cross section 
$\sigma (e^+e^- \to H) = |A_H|^2 \, k$ for the `heavy' channel and $\sigma
(e^+e^- \to L) = |A_L|^2 \, p_L$ for each of the `light' channels with $p_L$
being the c.m. momentum in the corresponding `light' channel at the `heavy'
threshold, which momentum is taken as a constant across the considered energy
range.

\begin{figure}[h]
\centering
\includegraphics[width=80mm]{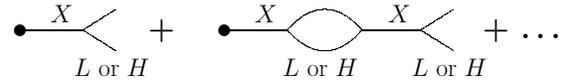}
    \caption{The graphs for the self energy of the resonance $X$ near the
threshold for the `heavy' channel ($H$). The absorptive and dispersive parts of
the self energy arise from loops with the `light' ($L$) and heavy meson pairs.
The filled circle stands for the electromagnetic vertex.}
\label{blobs}
\end{figure}

The expressions (\ref{ah}) and (\ref{al}) can be readily found from the standard
summation of the `blobs' in the graphs of the type shown in Figure~\ref{blobs},
where the ultraviolet part of the loop is absorbed into the definition of the
overall normalization and of the position $W_0$ of the resonance. However the
general treatment of the threshold behavior of the amplitudes\cite{ll}
guarantees that these expressions are quite independent of assumptions about the
form factors used in the graphs of Fig.~\ref{blobs}.

Furthermore, the `nominal' width $\Gamma_0$ is due to the coupling to the
lighter channels, and the term $g_H^2 \, k^3$ in the denominator in equations
(\ref{ah}) and (\ref{al}) represents the additional contribution to the
resonance width associated with the decay into the `heavy' meson pair, and $g_H$
is the coupling constant with the dimension of length. This extra contribution
to the width can be written in terms of the excitation energy $W $ as $g_H^2 \,
k^3 = W \, \sqrt{W/\epsilon}$, where
\beq
\epsilon=g_H^{-4} \, M^{-3}
\label{wp}
\eeq
is a parameter with dimension of energy.
The critical point of the present consideration is that the amplitudes $A_L$ as
functions of the energy $W$ are analytical functions of $W$ (with a cut at
positive $W$) and the expression (\ref{al}) can be analytically continued to
negative values of $W$. Using the notation $\Delta = - W$ at $W < 0$, the
amplitudes $A_L$ below the `heavy' threshold take the form
\beq
A_L=a_L + {b_L  \over {1 \over 2} \Delta \sqrt{\Delta \over \epsilon} - \Delta -
W_0 + {i \over 2} \,  \Gamma_0 }~~~.
\label{all}
\eeq

The term in the denominator in the latter formula, proportional to
$\Delta^{3/2}$, is clearly the analytical continuation of the $g_H^2 \, k^3$
term in Eq.~\ref{al}, and the proper sign is as indicated in
Eq.~\ref{all}~\footnote{This sign is due to the $P$ wave motion, and is opposite
to the one that would arise in a similar term for an $S$ wave ($\propto -
\Delta^{1/2}$).}. This term becomes more essential than the one linear in
$\Delta$, starting from $\Delta \approx \epsilon$, and gives rise to the
discussed channel coupling effects. In particular, the expression ${1 \over 2}
\Delta \sqrt{\Delta \over \epsilon} - \Delta$ cannot be arbitrarily negative and
reaches its minimum, equal to $-16 \, \epsilon/27$ at $\Delta = \Delta_m= 16 \,
\epsilon/9$. Thus in the case where the resonance is below the $D^* {\bar D}^*$
threshold, i.e. at negative $W_0=-\Delta_0$, the real part of the resonance
denominator has either two zeros if $\Delta_0 < 16 \epsilon/27$, or no zero at
all otherwise. In the latter case the position of the minimum of the absolute
value of the denominator, i.e. of the maximal effect on the cross section, is at
$\Delta=\Delta_m$ and does not depend on $\Delta_0$ at all, but rather is
determined by the parameter $\epsilon$. The energy corresponding to
$\Delta=\Delta_m$ in fact also determines the position of the maximum of the
resonance amplitude in the case where the `nominal' position of the resonance is
{\it above} the threshold, i.e. at positive $W_0$. This is due to the fact that
at sufficiently small $\epsilon$ the damping of the `light' channel amplitude
$A_L$ due to the absorptive term $g_H^2 \, k^3$ in the denominator in
Eq.~\ref{al} suppresses the `light' channel above the `heavy' threshold.

Clearly, the significance of the discussed threshold effect critically depends
on the value of the parameter $\epsilon$ for the near-threshold resonance. At
least for two known resonances, $\psi(3770)$ and $\Upsilon(4S)$, at the
thresholds of pairs of pseudoscalar heavy mesons the value of $\epsilon$ can be
found from the data: $\epsilon \approx 60\,$ MeV for $\psi(3770)$, and $\epsilon
\approx 20\,$ MeV for $\Upsilon(4S)$. There is every reason to expect, from the
observed onset of the $D^* {\bar D}^*$ production, that for the discussed
possible resonance the coupling to the vector meson pair should be stronger than
for the case of pseudoscalar mesons, and the parameter $\epsilon$ should be
significantly smaller than for $\psi(3770)$. The parameter $\epsilon$ is
proportional to $g_H^{-4}$ (cf. Eq.~\ref{wp}). Therefore, it is quite likely
that for the $D^* {\bar D}^*$ threshold resonance this parameter can be smaller
than that for the $\psi(3770)$ by a factor of 10 or more. It would come as no
surprise if the value of $\epsilon$ will be eventually found between 1 and 10
MeV.

\begin{figure*}[t]
\centering
\includegraphics[width=135mm]{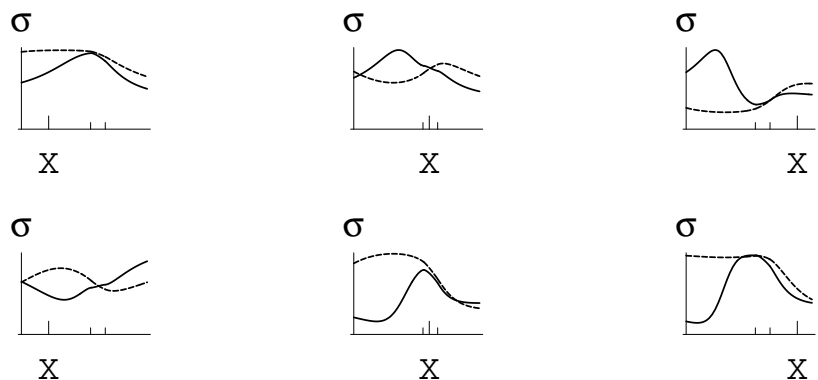}
    \caption{The types of behavior of the $e^+e^-$ annihilation cross section
(in arbitrary units) into one of the `light' channels near the $D^* {\bar D}^*$
threshold. The horizontal axis in each plot spans the c.m. energy range from
3980 MeV to 4040 MeV. The $D^{*0} {\bar D}^{*0}$ and $D^{*+} {\bar D}^{*-}$
thresholds are shown with shorter vertical tick marks. The plots are shown for
three assumed values of the `nominal' position $W_0$ of the resonance $X$
relative to the $D^{*0} {\bar D}^{*0}$ threshold: -20, 3, and 20 MeV, and the
assumed value for this position is indicated in each plot by the longer vertical
tick mark. In the plots in the upper row it is assumed that the coefficients
$a_L$ and $b_L$ are relatively real and have the same sign. In the lower row of
plots the relative sign of these coefficients is assumed to be negative. The
solid lines in the plots correspond to $\epsilon=1\,$MeV, and the dashed one are
for $\epsilon=10\,$MeV. The width parameter $\Gamma_0$ is fixed at
$\Gamma_0=50\,$MeV. } 
\label{xsections}
\end{figure*}

The possible types of behavior of the production cross section in a `light'
channel in the narrow energy range around the $D^* {\bar D}^*$ threshold is
illustrated in Figure~\ref{xsections} for some representative values of the
parameters, and where the isotopic splitting of the threshold is also taken into
account. It is worth mentioning that the interference of the resonant amplitude
with the non-resonant background is generally different for different channels,
which may result in a significantly different behavior of the observed shape of
the cross section for each channel. It is important to notice however that the
features in the behavior of the cross section generally do not coincide with the
position of the resonance. The shape of the cross section of the $e^+e^-$
annihilation into the vector  meson pairs $D^* {\bar D}^*$ above the threshold 
is only weakly sensitive to the parameters of the resonance and is dominated by
the very rapid rise from the threshold due to the $P$ wave factor $k^3$.

\section{Angular Correlations in $D^* {\bar D}^*$ Production Near the Threshold}

As previously mentioned the threshold resonance should couple to a specific
mixture of the states of the $D^* {\bar D}^*$ pair with the total spin $S=0$ and
$S=2$. The coefficients in this mixture and their relative phase can be studied
by measuring angular correlations in the production in $e^+e^-$ annihilation
near the threshold~\cite{ov43}. It would also be of interest to study the
behavior of these coefficients with increasing energy in order to possibly
untangle the resonant and the non-resonant production amplitudes.

Let $A_0$ and $A_2$ denote the production amplitudes for correspondingly the
$S=0$ and $S=2$ states, normalized in such a way that the total cross section is
given by $\sigma(e^+e^- \to D^* {\bar D}^*)=|A_0|^2 + |A_2|^2$. Then the
distribution of the production rate in the angle $\theta$ between the direction
of the momentum of either of the mesons and the $e^+e^-$ beams has the form
\beq
{{\rm d} \sigma \over {\rm d} \cos \theta} \propto 1-{|A_2|^2+10 \, |A_0|^2
\over 7 \, |A_2|^2 + 10 |A_0|^2} \, \cos^2 \theta~.
\eeq
Clearly, this angular distribution is not sensitive to the relative phase
$\varphi$ of the amplitudes $A_0$ and $A_2$. Such sensitivity however arises in
the distribution over the angle $\vartheta_\pi$ between the momentum of the
parent $D^*$ (${\bar D}^*$) and the $D$ (${\bar D}$) emerging from the decay
$D^* \to D \pi$ (${\bar D}^* \to {\bar D} \pi$), or the similar correlation over
the angle $\vartheta_\gamma$ for the events with the decays $D^* \to D \gamma$
(${\bar D}^* \to {\bar D} \gamma$). The angular distributions are expressed
through the two discussed amplitudes as
\begin{eqnarray}
{{\rm d} \sigma \over {\rm d} \cos \vartheta_\pi} \propto  |A_0|^2 + {1 \over
20}\, |A_2|^2 (13+21 \cos^2 \vartheta_\pi)\nonumber \\  +  {2 \over \sqrt{5}} \,
|A_0| \, |A_2| \, \cos \varphi \, (3 \, \cos^2 \vartheta_\pi- 1)~,
\end{eqnarray}
and
\begin{eqnarray}
{{\rm d} \sigma \over {\rm d} \cos \vartheta_\gamma} \propto |A_0|^2 + {1 \over
20}\, |A_2|^2 (13+{21 \over 2} \sin^2 \vartheta_\gamma) \nonumber \\
+ {2 \over \sqrt{5}} \, |A_0| \, |A_2| \, \cos \varphi \, ({3 \over 2}\, \sin^2
\vartheta_\gamma- 1)~.
\end{eqnarray}
The latter two angular distributions however are not independent: one is found
from the other by a simple substitution $\cos^2 \vartheta_\pi \leftrightarrow {1
\over 2} \, \sin^2 \vartheta_\gamma$. The reason of course is that they both
contain the same information about the vector meson polarization density matrix.
Thus measuring one or the other, or both, depends on the specific experimental
setup.

\section{Summary}

It is quite likely that in the resonance X(3872) we have come across a
`molecular' state made out of dominantly $D^0$ and $D^{*0}$. Although the
existence of such states for ``sufficiently heavy" heavy hadrons was not
unexpected theoretically, it was not clear whether the charmed mesons were heavy
enough. Given the present uncertainty in understanding the strong dynamics of
systems of heavy hadrons it is by far not clear yet whether the X(3872) is a
result of coincidence of several factors facilitating the binding, and no other
such states in the charmonium mass region exist, or there are other similar
states with hidden charm. One such candidate for a molecular state is the
apparent strong threshold singularity in $e^+e^-$ annihilation to $D^* {\bar
D}^*$. Further theoretical understanding of the strong interaction between heavy
mesons can be (hopefully) facilitated by a more detailed study of the resonance
X(3872) and by a detailed scan of the $e^+e^-$ annihilation into exclusive
channels with charmed mesons in the vicinity of the $D^* {\bar D}^*$ threshold.
In this respect the interesting studies which look feasible include the
measurement of the decays $X(3872) \to D^0 {\bar D}^0 \pi$,  $X(3872) \to D^0
{\bar D}^0 \gamma$ (the photon spectrum in this decay is directly related to the
motion of the mesons inside the X resonance), a search for the decay $X(3872)
\to D^+D^- \gamma$, which according to the presented discussion is sensitive to
the transition $X(3872) \to \psi(3770) \gamma$, a search for the decays $X(3872)
\to \chi_{cJ} \pi^0$. For understanding the structure of the resonance very near
the $D^* {\bar D}^*$ threshold in the $e^+e^-$ annihilation it is most
interesting to have a detailed scan of the production cross section for the
lighter pairs: $D {\bar D}$, $D^* {\bar D}$ and $D_s {\bar D}_s$, since these
channels may exhibit a quite intricate behavior sensitive to detailed parameters
of the resonance. Also a study of the angular correlations in $e^+e^- \to D^*
{\bar D}^*$ near the threshold would provide an information on the `hyperfine'
structure of the possible molecular state.

\bigskip 
\begin{acknowledgments}
I thank Brian Lang for discussions of the angular correlations in the $D^* {\bar
D}^*$ production in $e^+e^-$ annihilation.
This work is supported in part by the DOE grant DE-FG02-94ER40823.
\end{acknowledgments}

\bigskip 

\end{document}